\documentclass[aps,prl,twocolumn,superscriptaddress,amssymb,amsmath,showpacs,floatfix]{revtex4-1}

\usepackage{graphicx}
\usepackage{dcolumn}
\usepackage{bm}
\usepackage{amsmath}
\usepackage{xfrac}

\begin{document}

\title{Energy limitation of laser-plasma electron accelerators}

\author{D.E. Cardenas}
\affiliation{\scriptsize Max-Planck-Institut f{\"u}r Quantenoptik, Hans-Kopfermann Strasse 1, 85748, Garching, Germany}
\affiliation{\scriptsize Ludwig-Maximilian-Universit{\"a}t M{\"u}nchen, Am Couloumbwall 1, 85748, Garching, Germany}
\author{S. Chou}
\affiliation{\scriptsize Max-Planck-Institut f{\"u}r Quantenoptik, Hans-Kopfermann Strasse 1, 85748, Garching, Germany}
\affiliation{\scriptsize Ludwig-Maximilian-Universit{\"a}t M{\"u}nchen, Am Couloumbwall 1, 85748, Garching, Germany}
\author{J. Xu}
\affiliation{\scriptsize Max-Planck-Institut f{\"u}r Quantenoptik, Hans-Kopfermann Strasse 1, 85748, Garching, Germany}
\affiliation{\scriptsize State Key Laboratory of High Field Laser Physics, Shangai Institute of Optics and Fine Mechanics, Chinese Academy of Sciences, P.O. Box 800-211, Shanghai, China}
\author{L. Hofmann}
\affiliation{\scriptsize Max-Planck-Institut f{\"u}r Quantenoptik, Hans-Kopfermann Strasse 1, 85748, Garching, Germany}
\affiliation{\scriptsize Ludwig-Maximilian-Universit{\"a}t M{\"u}nchen, Am Couloumbwall 1, 85748, Garching, Germany}
\author{A. Buck}
\affiliation{\scriptsize Max-Planck-Institut f{\"u}r Quantenoptik, Hans-Kopfermann Strasse 1, 85748, Garching, Germany}
\affiliation{\scriptsize Ludwig-Maximilian-Universit{\"a}t M{\"u}nchen, Am Couloumbwall 1, 85748, Garching, Germany}
\author{K. Schmid}
\affiliation{\scriptsize Max-Planck-Institut f{\"u}r Quantenoptik, Hans-Kopfermann Strasse 1, 85748, Garching, Germany}
\affiliation{\scriptsize Ludwig-Maximilian-Universit{\"a}t M{\"u}nchen, Am Couloumbwall 1, 85748, Garching, Germany}
\author{C.M.S. Sears}
\affiliation{\scriptsize Max-Planck-Institut f{\"u}r Quantenoptik, Hans-Kopfermann Strasse 1, 85748, Garching, Germany}
\affiliation{\scriptsize Ludwig-Maximilian-Universit{\"a}t M{\"u}nchen, Am Couloumbwall 1, 85748, Garching, Germany}
\author{D.E. Rivas}
\affiliation{\scriptsize Max-Planck-Institut f{\"u}r Quantenoptik, Hans-Kopfermann Strasse 1, 85748, Garching, Germany}
\affiliation{\scriptsize Ludwig-Maximilian-Universit{\"a}t M{\"u}nchen, Am Couloumbwall 1, 85748, Garching, Germany}
\author{B. Shen}
\affiliation{\scriptsize State Key Laboratory of High Field Laser Physics, Shangai Institute of Optics and Fine Mechanics, Chinese Academy of Sciences, P.O. Box 800-211, Shanghai, China}
\author{L. Veisz}
\affiliation{\scriptsize Max-Planck-Institut f{\"u}r Quantenoptik, Hans-Kopfermann Strasse 1, 85748, Garching, Germany}

\date{\today}

\begin{abstract}

\indent We report on systematic and high-precision measurements of dephasing, an effect that fundamentally limits the performance of laser wakefield accelerators. Utilizing shock-front injection, a technique providing stable, tunable and high-quality electron bunches, acceleration and deceleration of few-MeV quasi-monoenergetic beams were measured with sub-5-fs and 8-fs laser pulses. Typical density dependent electron energy evolution with 65-300 $\mu$m dephasing length and 6-20 MeV peak energy was observed and is well described with a simple model.

\end{abstract}


\maketitle

\indent Laser wakefield acceleration (LWFA), a laser-driven electron acceleration scenario, was proposed in 1979 \cite{Tajima1979} and realized first in the tens of MeV energy range, \cite{Geddes2004,Faure2006,Leemans2006,Gonsalves2011} and later at lower \cite{Schmid2009} as well as higher energies \cite{Kim2013,Wang2013,Leemans2014}, generating multi-GeV beams. LWFA offers longitudinal accelerating fields on the order of 100 GV/m, many orders of magnitude larger than conventional accelerators and it enables the generation of few-fs pulse duration electron bunches \cite{Buck2011,Lundh2011}. Therefore, compactness, novelty and accessibility for possible future applications. The maximum electron energy in a single accelerator stage is limited by laser diffraction, laser depletion and by the fact that the electrons are eventually faster than the laser and the plasma wave. The latter is referred to as electron dephasing and the corresponding maximum acceleration length is the dephasing length $L_{d}$. It is given approximately by $L_{d,\mathrm{1D}} \approx \lambda_{p}^3/\lambda_{0}^2$ for underdense plasmas in the 1D nonlinear theory \cite{Esarey1996}, where $\lambda_{p} \,\left [\mathrm{\mu m} ] \right. =3.3\times10^{10}/\sqrt{n_e \,\left [ \mathrm{cm^{-3}} ] \right.}$ and $\lambda_{0}$ are the plasma and laser wavelengths, respectively, and $n_e$ is the electron density \cite{Esarey2009}. For acceleration lengths larger than the dephasing length, the electron bunch starts to decelerate as the electric field within a plasma period has changed sign. On the other hand, for acceleration lengths shorter than the dephasing length, the electron bunch always experiences an accelerating field and thus, its energy increases. The energy increase continues until the longitudinal electric field reaches zero, at which point the acceleration length reaches $L_d$ and the maximum energy is obtained. In order to reach higher energies $>\mathrm{GeV}$, all energy limitations must be controlled. Various techniques have been implemented to mitigate diffraction such as preformed plasma channels \cite{Geddes2005}, gas-filled capillary discharge waveguides \cite{Leemans2006} or laser self-channeling \cite{Mangles2012}. In the high energy experiments cited above, in order to reach $\mathrm{GeV}$ range, a relatively low ($\approx 10^{17}-10^{18} \,\mathrm{cm^{-3}}$) electron density was used, thus electron dephasing was not the primary limiting factor. Instead, the laser was kept focused as long as possible, so that the electrons stayed in phase and were longer accelerated. Even if the electron dephasing is prevented by using low electron densities, there are still proposals to enlarge the dephasing length by using a tapered plasma medium with increasing density \cite{Rittershofer2010,Yu2014}. Apart from the energy limitation, it has been simulated that the dephasing effect improves the beam quality. Therefore, better control and understanding of this effect is fundamental to the physics of laser plasma accelerators.

\indent In order to resonantly excite a plasma wave, the laser pulse length should match half of the plasma wavelength, i.e. $c\tau_{\mathrm{pulse}} = \lambda_{p}/2$ \cite{Tajima1979}. Up to now, ultrashort ($\sim\,25\,\mathrm{fs}$) highly intense ($I>10^{18} \,\mathrm{Wcm^{-2}}$, $a_0>1$, where $a_0$ is the normalized laser vector potential) Ti:Sapphire laser pulses ($\lambda_0 = 800 \mathrm{nm}$) were typically employed, which led to dephasing lengths of many mm's. This represents a difficulty in terms of measurability as long gas targets are needed and electron bunches with quasi-monoenergetic spectra and narrow energy spread. As $L_d \varpropto \lambda_{p}^3 \varpropto \tau_{\mathrm{pulse}}^3$, shorter pulses, matched densities and relativistic intensities provide a measurable dephasing effect. Optical Parametric Chirped Pulse Amplification (OPCPA) \cite{Dubietis1992} technology allows the production of multi-TW sub-10 fs laser pulses as demonstrated by the Light Wave Synthesizer 20 (LWS-20) at the Max Planck Institute of Quantum Optics: LWS-20 older version (8 fs) \cite{Herrmann2009} was and its upgrade, ($< 5 \,\mathrm{fs}$) \cite{Veisz2014}, is the most intense few-cycle laser system in the world. Pulses as short as these give rise to an acceleration which dephases within about 65-300 $\mu$m, and thus it is measurable for the first time with high precision. 

\indent In this paper, we present the first direct and systematic measurement of the electron dephasing in a laser wakefield accelerator. To this end, the acceleration length is scanned by varying the position of electron injection into the plasma wave. The injection mechanism used in this experiment is the so called shock-front injection \cite{Schmid2010}. It has proved to work for long pulses ($\sim 25\,\mathrm{fs}$) \cite{Buck2013} as well as for short ones $\sim 5\,\mathrm{fs}$ providing stable high quality quasi-monoenergetic electron beams with low energy spread ($\Delta E_{\mathrm{FWHM}}<5$ MeV) and tunability of the electron spectrum over more than one order of magnitude by using a long laser. Shock-front injection is realized by placing a razor blade on top of a supersonic gas nozzle, creating an ultra-thin ($< 5\, \mathrm{\mu m}$) shock-front which separates two different plasma regions with a certain density ratio ($\approx 1.6$ in our case, $n_{1}>n_{2} \rightarrow \lambda_{p,1}<\lambda_{p,2}$). As the laser passes the density transition, a fraction of the background plasma electrons, forming the end of the first plasma period in the high density range, are injected in the accelerating phase of the first plasma period in the lower density range. Such a sharp density-transition allows an instantaneous and precise injection and therefore, very low and energy independent absolute energy spread of the electron beam. This simple but effective method provides high control over the electron injection position and thus over the final peak energy, by shifting the blade and correspondingly the shock-front position.

\begin{figure}[ht!]
\centering
\includegraphics[width=0.9\linewidth]{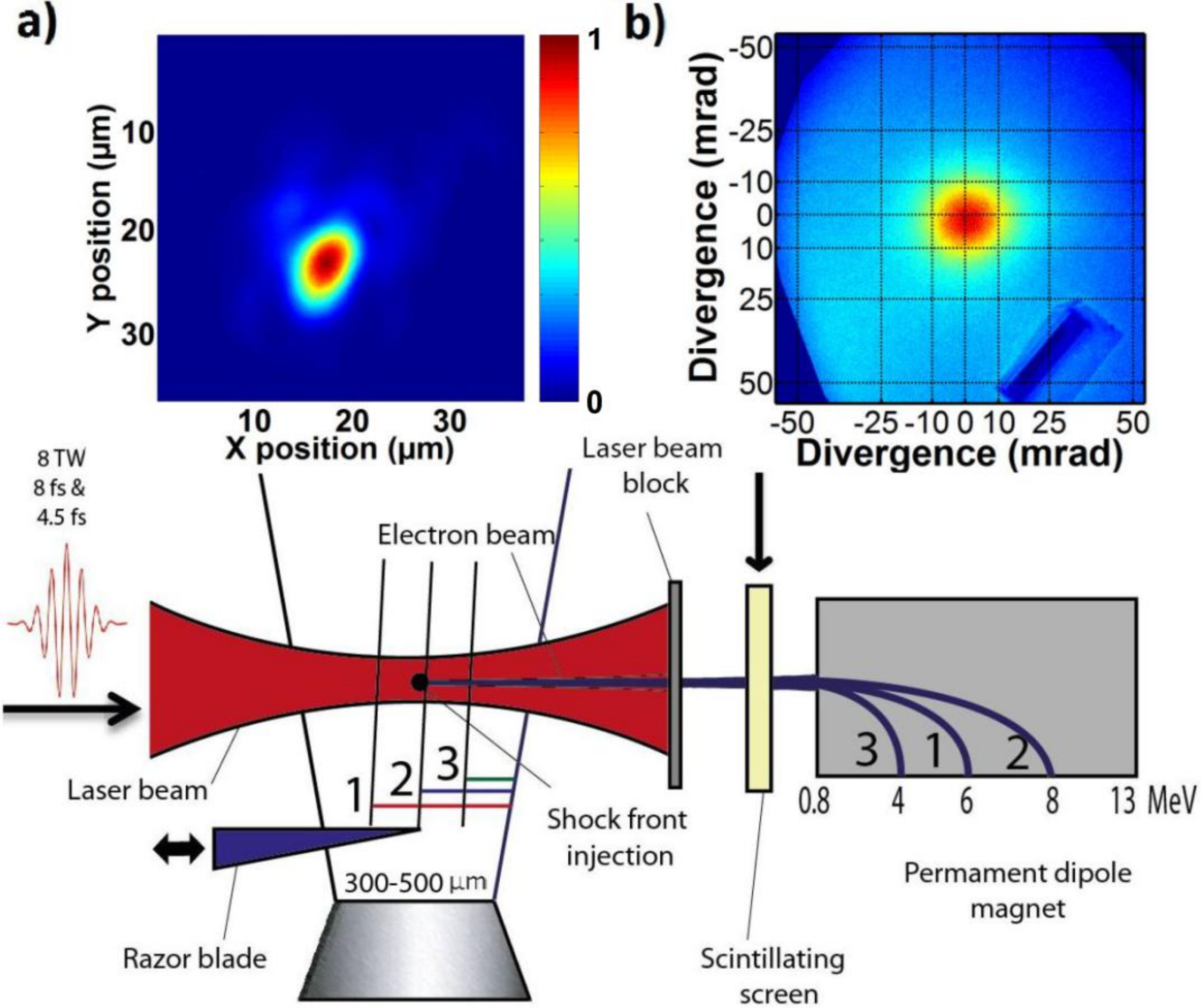}
\caption{Experimental setup: The laser beam was focused a $300-500 \,\mu$m supersonic de Laval nozzle. A razor blade, which produced the shock front, was moved longitudinally to tune the  acceleration length of the injected electron bunch was tuned. A removable scintillating screen (BIOMAX) was used for absolute charge measurement and beam profile observation. A permanent dipole magnet, equipped with BIOMAX as well, was installed to measure the electron spectrum. Different acceleration lengths are illustrated: longer (1), similar (2) and shorter (3) than the dephasing length. Typical electron spectrum from (1),(2) and (3) are shown in Fig.\,\ref{typical_spectrum_5fs}. Inset: Typical laser focal spot (a) and electron beam profile (b) in vacuum in the sub-5-fs case: 1 pC charge and 10-mrad root-mean-square (rms) divergence. Both insets have the same color scale.}
\label{setup}
\end{figure}

\indent In our experiments (Fig.\,\ref{setup}), we focused the old version of LWS-20 with 8-fs, 800 nm central wavelength, 130 mJ pulses ($65\, \mathrm{mJ}$ on target) to a spot of $12 \,\mu$m ($\mathrm{FWHM}$), and the LWS-20 with 4.5 fs, 740 nm, 80 mJ pulses ($37 \,\mathrm{mJ}$ on target) to a spot of $5.5 \,\mu$m ($\mathrm{FWHM}$), reaching peak intensities of $2.8 \times 10^{18} \,\mathrm{W\,cm^{-2}}$ and $1.2 \times 10^{19}\, \mathrm{W\, cm^{-2}}$, respectively. A 300 $\mu$m (and 500 $\mu$m for the 8-fs case) supersonic de Laval nozzle with helium gas and a razor blade on a translation stage provided the acceleration medium with controlled injection. The electron density was tuned within the range of $4-21 \times 10^{19} \,\mathrm{cm^{-3}}$ to approximately match the corresponding laser pulse duration. Shadowgraphy with a probe beam propagating perpendicularly to the main beam was used to observe the plasma channel, plasma wave and the injection position of the electrons. The plasma wavelength, i.e. the electron density, is matched to the laser pulse duration in a pure LWFA scenario. However, within a small range of densities away from this resonance, wakefield excitation and electron acceleration also take place. This enabled us to measure the dephasing effect in more than two cases, even though only two different pulse durations were employed.

\begin{figure}[ht!]
\centering
\includegraphics[width=0.9\linewidth]{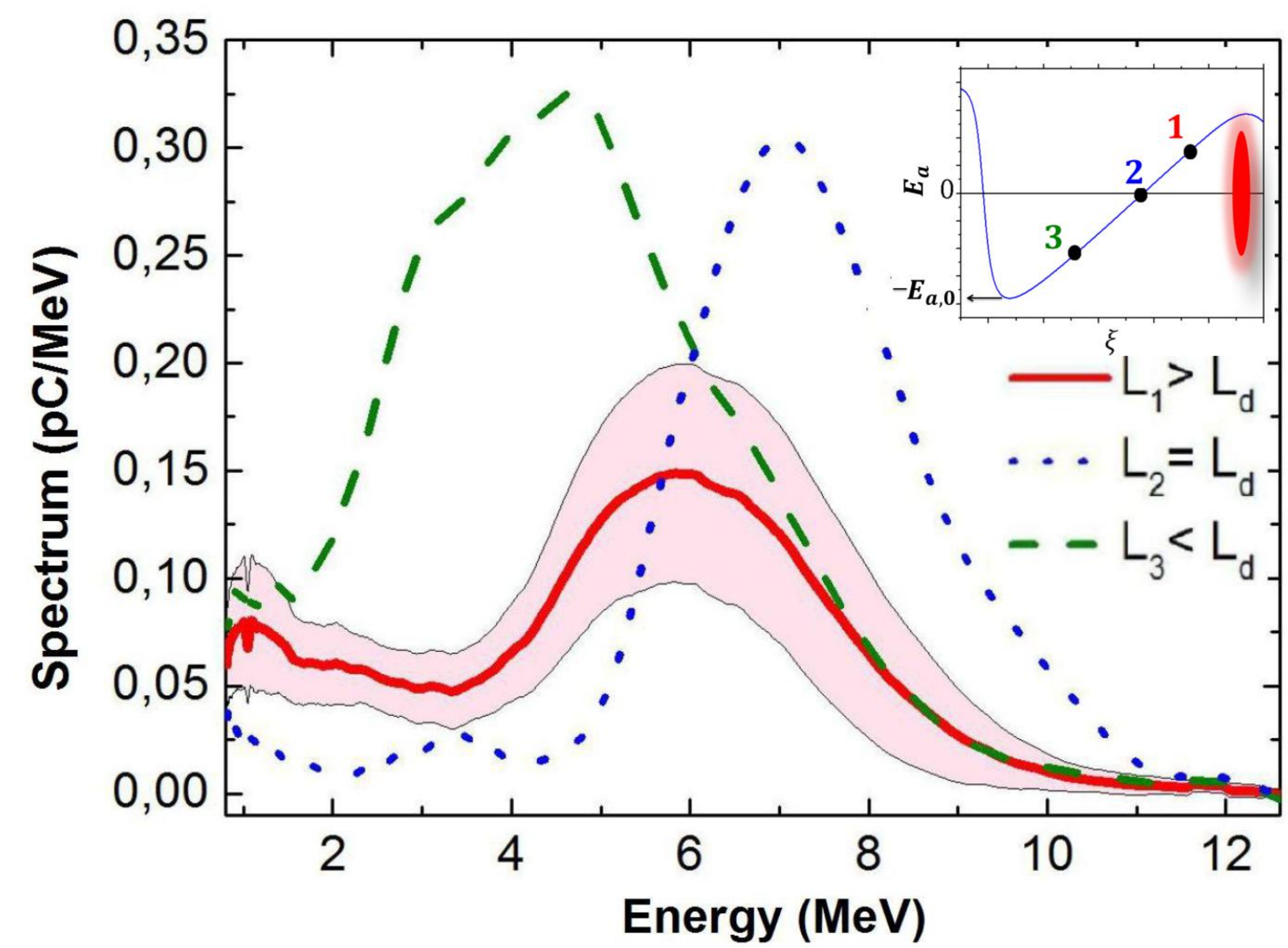}
\caption{Average electron spectrum over 20 shots for ($L_1$) and standard deviation band in shaded area, and two typical single-shot electron spectra (for $L_2 \,\, \mathrm{and} \,\, L_3$) using $7.7 \times 10^{19}\,\mathrm{cm^{-3}}$ and 4.5-fs LWS-20 pulses. The estimated dephasing length is of $L_d \approx 130 \, \mu\mathrm{m}$. Inset: Electric field of the plasma and position of the electrons in the co-moving frame inside the first plasma period at the time they leave the plasma. Electrons 1,2 and 3 have $L_{1}=200\, \mu\mathrm{m},L_{2}=150\,\mu\mathrm{m}$ and $L_{3}=50 \,\mu\mathrm{m}$ acceleration lengths, respectively.
See also Fig.\,\ref{setup}.}
\label{typical_spectrum_5fs}
\end{figure}

\begin{figure}[ht!]
\centering
\includegraphics[width=0.9\linewidth]{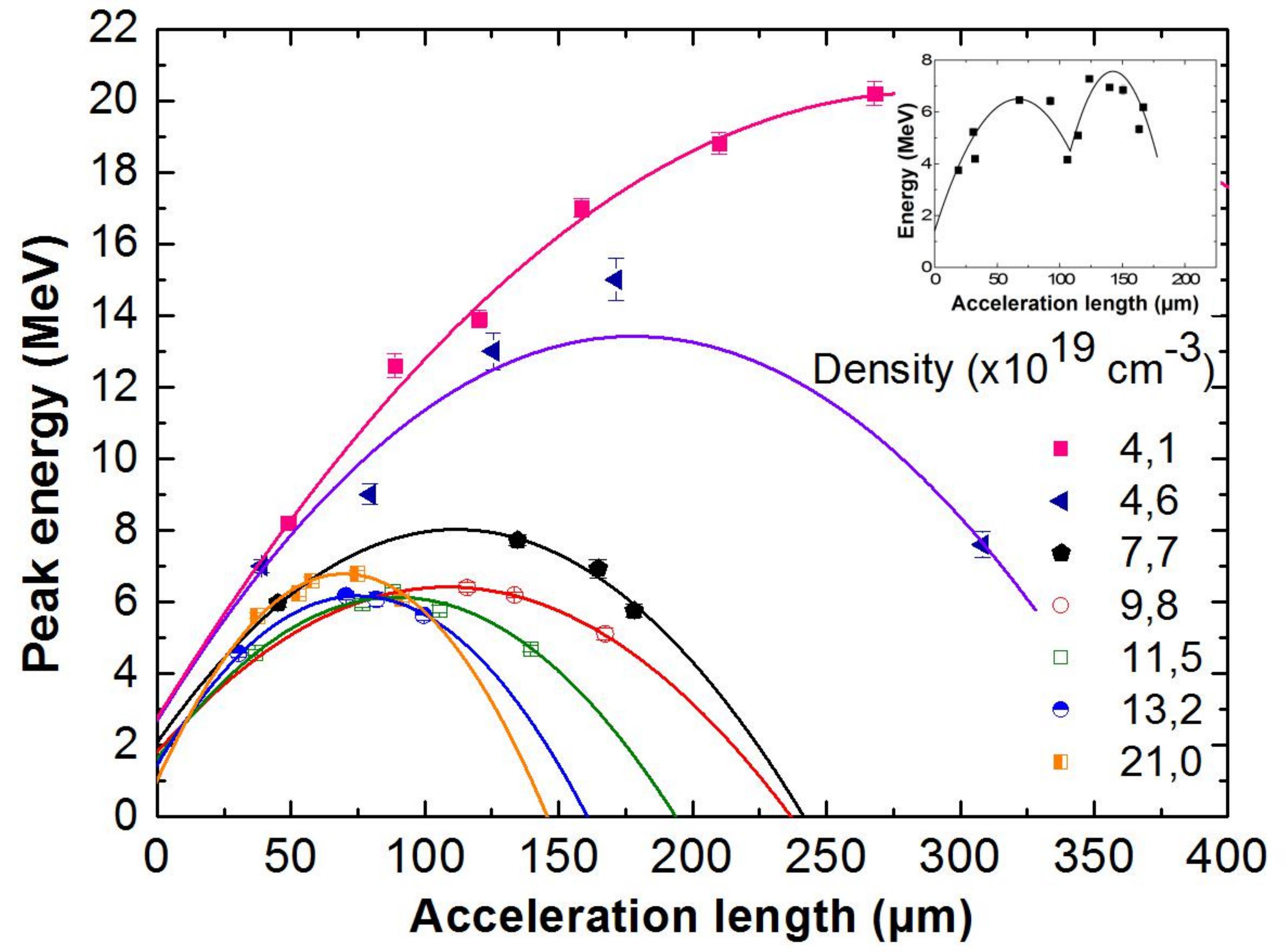}
\caption{Peak energy of the electron spectrum, after subtraction of plasma background electron energy, for sub-10-fs laser pulses vs. acceleration lengths for various electron densities. The errorbars indicate standard error over 50 shots. The lines are fits using Eq.\,\eqref{fit}. Inset: Peak energy of the electron spectra also much beyond dephasing at a density of $14.4 \times 10^{19}\,\mathrm{cm^{-3}}$ ($L_a>L_d$). The fit from Eq.\,\eqref{fit} for the first and separately second period of dephasing are also plotted.}
\label{curves}
\end{figure}

\indent For the 8-fs case, typical electron bunches had approximately $4 \,\mathrm{pC}$ charge, peak energies of 5-20 $\mathrm{MeV}$ and approximately 4 mrad rms divergence. While for the sub-5 fs-case, 1 pC bunches with rather large divergence (10 mrad rms) bunches were generated at $<10\, \mathrm{MeV}$ peak energies. The electron spectra are plotted for different acceleration lengths in Fig.\,2 using sub-5-fs laser pulses. The electron energy increased with acceleration length $(L_3)$, until this distance reached approximately the dephasing length $(L_2)$. Here the energy was maximal and started to decrease afterward $(L_1$). The degree of dephasing is well illustrated by the position of the electron bunch in the accelerating electric field in the co-moving frame by the time the plasma ends. This is shown in the inset in Fig.\,\ref{typical_spectrum_5fs}, using a 1D model \cite{Gibbon} for the nonlinear plasma electric field. The acceleration lengths were scanned in a range large enough for the dephasing effect to manifest. This procedure was repeated for electron densities of $4.1 \times 10^{19}\,\mathrm{cm^{-3}}$ and $4.6 \times 10^{19}\,\mathrm{cm^{-3}}$ for the older 8-fs LWS-20 and from $7.7 \times 10^{19}\,\mathrm{cm^{-3}}$ up to $21 \times 10^{19}\,\mathrm{cm^{-3}}$ with the sub-5-fs system. The observed peak energies are plotted in Fig.\,\ref{curves}. To also exclude laser depletion, the laser energy was measured after acceleration and a loss of only 10-20$\%$ was obtained. The relative energy spread decreased strongly for short acceleration lengths ($<L_d$) and became approximately constant when dephasing was reached. The divergence of the electron beam during the whole dephasing period (acceleration and deceleration) scaled as $1/\gamma_e$, where $\gamma_e$ is the gamma factor of the electrons.

\indent A simple model is presented to obtain the dephasing length from the experimental results. In a nonlinear plasma wave, the longitudinal electric field in the first plasma period scales roughly linearly with the position in the co-moving frame $\xi$, as seen in the inset in Fig.\,\ref{typical_spectrum_5fs}. To the first order, this field also scales linearly within the laboratory frame during the first dephasing period. The longitudinal electric field of the plasma at the position of the electron bunch $E_{\mathrm{a}}$ as a function of its position in the laboratory system $x$ is:

\begin{equation}
\label{accfield}
E_{\mathrm{a}}(x) = E_{\mathrm{a},0}-\frac{E_{\mathrm{a},0}}{L_{\mathrm{d}}}x ;\,\, 0 \leq x \leq 2L_{\mathrm{d}}.
\end{equation}

where $E_{\mathrm{a},0}$ is the maximum accelerating field at the rear of the first plasma period. The accelerating field is zero just where the dephasing limit is reached. This simple model neglects beamloading and electron beam interaction with the laser. The electron energy is obtained by adding the integral of the field Eq.\,(\ref{accfield}) from the injection position $x = 0$ up to a certain acceleration length $x = L_{\mathrm{a}}$ and the electron energy corresponding to the velocity of the plasma wave $E_b= 0.511 \,\mathrm{MeV} (\lambda_p / \lambda_0 -1)$, which is the initial energy of a trapped electron. This yields 

\begin{align}
\label{fit}
E_{\mathrm{p}}(L_{\mathrm{a}}) & = \int\limits_{0}^{L_{\mathrm{a}}} E_{\mathrm{a}}(x)\, dx + E_{\mathrm{b}}\\ 
& = L_{\mathrm{a}} E_{\mathrm{a},0} (1-L_{\mathrm{a}}/2L_{\mathrm{d}})+E_{\mathrm{b}}, \nonumber
\end{align}

\begin{figure}[ht!]
\centering
\includegraphics[width=0.9\linewidth]{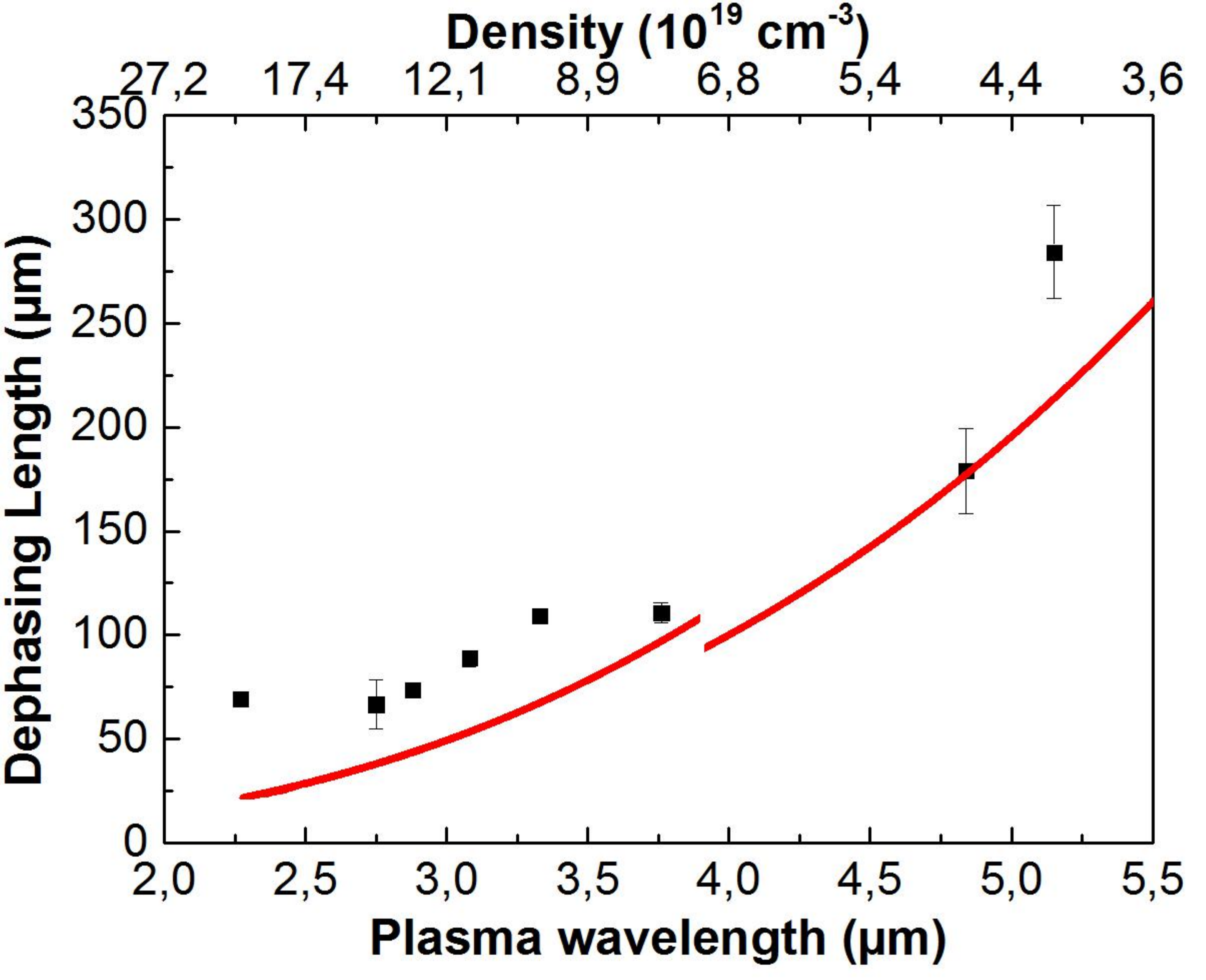}
\caption{Measured (squares) and predicted dephasing lengths $L_{d,\mathrm{1D}}$ (continuous line) for different electron densities. Electron densities above $7 \times 10^{19}\,\mathrm{cm^{-3}}$ were used only with more intense ($a_0 \approx 2.2$) sub-5-fs pulses, 740 nm central wavelength, while 8-fs pulses have $a_0 \approx 1.1$ and 800 nm.}
\label{deph_lengths}
\end{figure}

\indent The peak energy of the electron spectrum  $E_{\mathrm{p}}$ depends quadratically on the acceleration length $L_{\mathrm{a}}$, as shown above. From the fit coefficients, the maximum accelerating field as well as the dephasing length are retrievable. Our model is valid in a broad range of densities and correspondingly maximum electron energies. For lower densities, $L_{\mathrm{d}} E_{\mathrm{a},0} \gg E_{\mathrm{b}}$ applies. However, in this low energy regime, the phase velocity of the plasma is not so low (1-3 MeV) compared to the maximum obtained energy (6-20 MeV) and therefore this last term on the right side of Eq.\,\eqref{fit} is not negligible. Fig.\,\ref{curves} shows the fit given by Eq.\,\eqref{fit} to the measured electron peak energies as a function of the acceleration length for different densities. Our model describes well the relevant processes even beyond the dephasing length. We observed in many cases that for long enough acceleration lengths, $L_{\mathrm{a}}>L_{\mathrm{d}}$, the low energetic dephased electrons increased their energy again by initiating another dephasing period, as plotted in Fig.\,\ref{curves} (inset) for $14.4 \times 10^{19}\,\mathrm{cm^{-3}}$, contrary to the results in \cite{Corde2013}. This is a further proof that both the laser depletion and diffraction were negligible. The measured dephasing lengths are obtained from the fit for each density and are plotted in Fig.\,\ref{deph_lengths}. Our final result matches very well with the 1D nonlinear dephasing formula shown in the introduction, as seen in Fig.\,\ref{deph_lengths}. Such short dephasing lengths ($\sim65-300\,\mathrm{\mu m}$) are a direct consequence of employing ($<10 \,\mathrm{fs}$) pulses in matched electron density plasmas to accelerate electrons. On the other hand, these results also show a clear indication that in order to reach high energy electrons, longer pulses should be applied to excite the wakefield, i.e. $E_{\mathrm{p,max}} \propto L_{\mathrm{d}} E_{\mathrm{a},0} \propto \lambda_{p}^2 \propto \tau_{\mathrm{pulse}}^2$, as expected in the normal laser wakefield regime. This energy dependence with pulse duration was observed in our experiment by taking into account the ratio of electron peak energy in the resonant-density case for the two laser systems, i.e. $20\, \mathrm{MeV}/6 \, \mathrm{MeV} \approx (8^2/4.5^2)$ in Fig.\,\ref{curves}. As laser energy depletion was small after the interaction and even a second period of dephasing was observed many times, the energy limitation of the accelerated electrons in these experiments was clearly due to the dephasing in the wakefield, and not laser depletion nor diffraction. 

\indent The maximum accelerating field from the fit was on the order of $\sim 100-200\,\mathrm{GV/m}$, which was smaller than the theoretical value of up to 1 $\mathrm{TV/m}$ in this high electron density regime. This result was explained by the fact that a normalized laser vector potential of $a_{0}\ge 4$ is needed to reach full blow-out or the so-called bubble \cite{Pukhov2002,Jansen2014} and the applied values are $a_{0} \approx 1-2$. Furthermore, the pulses were not significantly self-focused due to their extremely short durations \cite{Sprangle2011} and could not self-compress \cite{Faure2005,Schreiber2010} due to dispersion in the plasma \cite{Beaurepaire2014}, especially in the sub-5-fs case. 

\begin{figure}[ht!]
\centering
\includegraphics[width=0.9\linewidth]{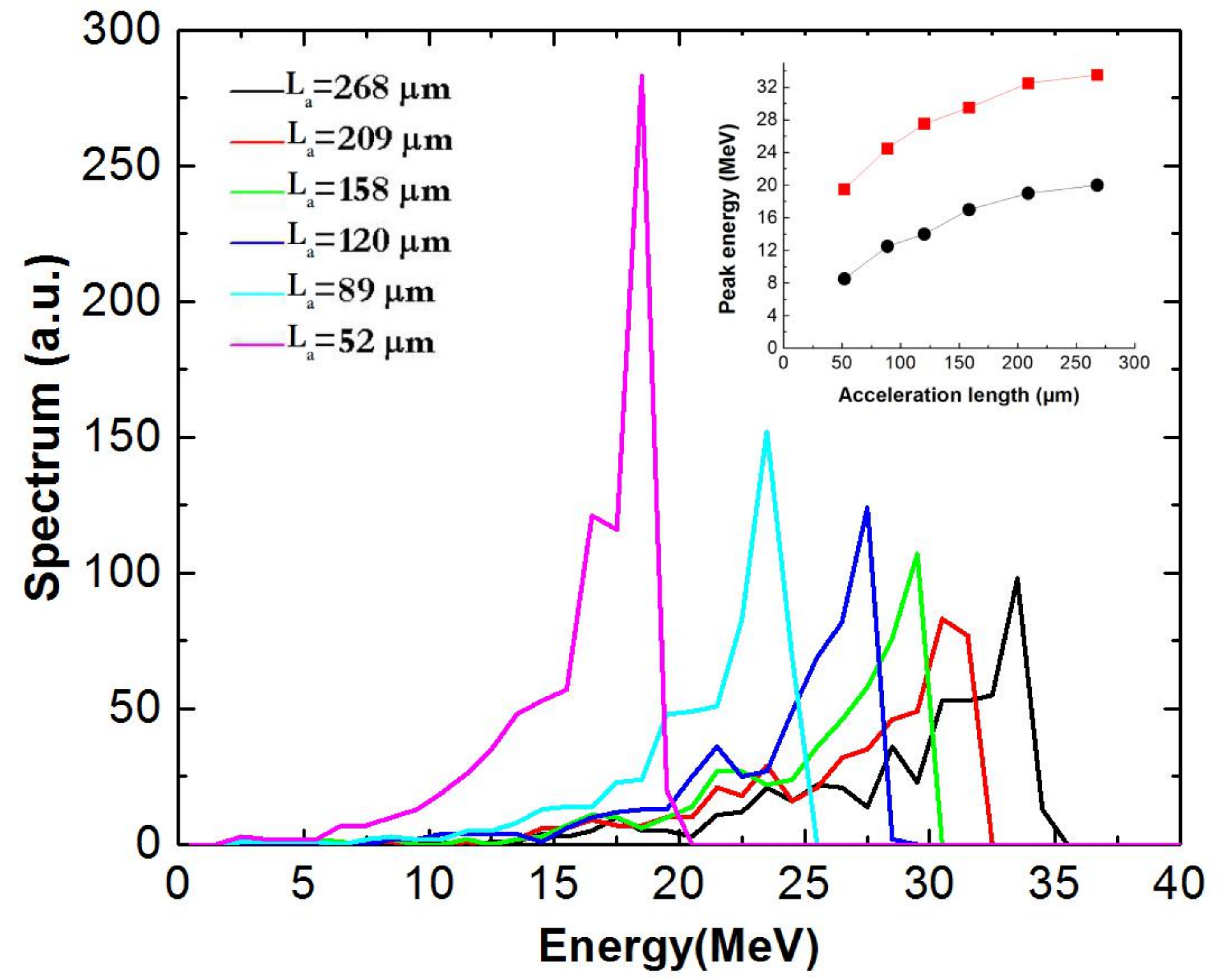}
\caption{Simulated electron spectra for 8-fs pulses for different acceleration lengths. Inset: Measured (circles) in the 8-fs case using $4.1 \times 10^{19}\,\mathrm{cm^{-3}}$ and 3D PIC simulation predicted (squares) electron peak energies as a function of the acceleration length.}
\label{8_fs_simu}
\end{figure}

3D particle-in-cell simulations were performed using the code VORPAL \cite{Nieter2004} to investigate details of the dephasing process. The simulation box was $20\times30\times30\,\mathrm{\mu m}^3$ and moved at the speed of light. It was divided into $400\times300\times300$ cells, each one containing one macro particle inside. These results agree qualitatively with the experimental observations and show similar charge, electron spectrum and dephasing length, as shown in Fig.\,\ref{8_fs_simu} for comparable conditions to the experimental values ($n_e = 1.8 \times 10^{19}\,\mathrm{cm^{-3}}$, 8-fs pulses and $a_0 \approx 0.75$). The following points were learned from the simulations: electrons are not trapped at the injection position at the shock front. After intense dynamics in the first few $\mu$m, a significant portion of the electrons leave the first period and most of the particles that will be trapped fall back almost to the end of the first plasma period before trapping. Therefore, the injection process will not shorten the dephasing length. Furthermore, there is a reduced accelerating field in the plasma wave due to not completely empty plasma wave cavity that originates from the modest driving laser intensities. 

\indent In conclusion, a systematic and direct measurement of the dephasing effect in a laser wakefield accelerator has been performed utilizing shock-front injection and sub-10-fs laser pulses. Quasi-monoenergetic electron beams have been generated at rather low relativistic energies due to rapid dephasing of the electrons in a high electron density plasma. $65-300\,\mu \mathrm{m}$ dephasing lengths have been measured by observing the electron energy as a function of the acceleration length changed by controlling the electron injection position along the plasma. These results suggest that shock-front injection is a highly reliable and useful technique in laser-plasma accelerators and allows, among others, systematic and precise characterization of dephasing. The study and control of the dephasing effect in laser accelerators is of primary importance to obtain the highest electron energy. Moreover, using such intense and short laser pulses as delivered by LWS-20, stable $<10 \,\mathrm{MeV}$ quasi-monoenergetic electron beams are generated. This is a promising electron source for ultrafast electron diffraction \cite{He2013} with sub-10-fs temporal resolution, once other challenges such as divergence and energy spread are overcome. 
\newline
\begin{acknowledgments}
\indent The authors thank Matthew Weidman for helpful comments on the manuscript. This work is supported by DFG-Project Transregio
TR18, The Munich Centre for Advanced Photonics (MAP) and by the Euratom research and training programme 2014-2018 under grant agreement No 633053 within the framework of the EUROfusion Consortium. 
\end{acknowledgments}

\bibliography{bibliography}

\end{document}